\documentclass[aps,prb,twocolumn,groupedaddress,floatfix,amssymb]{revtex4}

\usepackage{graphicx}% Include figure files
\usepackage{bm}% bold math
\usepackage{amsmath}% \text{}
\begin{document}

\title{Strongly Interacting Luttinger Liquid and Superconductivity in an Exactly Solvable Model}

\author{Igor N. Karnaukhov ${}^{1}$}
%\email[e-mail: ]{antonov@ameslab.gov; anton@imp.kiev.ua}
%\thanks{}
\altaffiliation[Permanent address: ]{Institute of Metal Physics, Vernadsky Street 36, 03142 Kiev, Ukraine}
%\affiliation{Max-Planck Institut f\"ur Physik Complexer Systeme, N\"othnitzer Stra\ss e 38, D-01187 Dresden, Germany}
\author{Alexander A. Ovchinnikov${}^{1,2}$}
\affiliation{${}^{1}$Max-Planck Institute f\"ur Physik Komplexer Systeme, N\"othnitzer Stra\ss e 38, D-01187 Dresden, Germany}
\affiliation{${}^{2}$Joint Institute of Chemical Physics, Kosygin Street 4, 117334 Moscow, Russia}

\date{\today}

%%%\tighten

%%%\twocolumn[\hsize\textwidth\columnwidth\hsize\csname@twocolumnfalse\endcsname

\begin{abstract}
A new family of exactly solvable one dimensional models with a hard-core repulsive potential
is solved by the Bethe Ansatz for an arbitrary hard-core radius. The exact ground state
phase diagrams in a plane 'electron density - on-site interaction' have been studied for
several values of a hard-core radius. It is
shown that superconducting phase and strongly interacting Luttinger liquid
state are coexisted at a high electron density and unusually high value
of repulsive on-site Coulomb interaction.
\end{abstract}
%%%\draft
%%%\preprint{IP/1}

\pacs{PACS numbers: 71.10.Fd; 71.10.Pm}
%\keywords{}
%%%] %\narrowtext
%\twocolumn


\maketitle

The models of strongly correlated electrons with a bond-charge interaction
which conserves a number of double occupied sites, are simple
examples of strongly correlated electron systems that exhibit
superconductivity \cite{EKS},\cite{AAS}, \cite{DM}. The merit of these
models is their complete integrability. The phase diagrams on a plane 'electron
density - on-site interaction' have four phases, two of them
exhibit off-diagonal long-range order
(ODLRO) and thus are superconducting. The superconducting phase is realized
if the value of the on-site interaction less then the critical one $U_{c}$ and
$U_c$ could be even positive at repulsive on-site interaction $U$. Other phases in which only
single occupied and empty sites are presented called as $U\rightarrow \infty $ Hubbard state.
In the case of a hard-core repulsive interaction between electrons with a hard core radius
which exceeds a half of a lattice spacing the Luttinger liquid state transforms
to strongly interacting Luttinger liquid in a high electron density region
\cite{IK}. Strongly interacting Luttinger liquid is characterized by a large
value of the critical exponent $\Theta$ for the momentum distribution function.
At  $\Theta >1 $ the residual Fermi surface disappears. At a high
electron density when a hard-core repulsion interaction dominates,
superconducting phase and strongly interacting
Luttinger liquid state do not coexist in the framework of the generalized
t-J and Lai-Sutherland models \cite{IK}. The question arises: could
the superconducting phase and strongly interacting Luttinger liquid state coexist simultaneously
at positive and finite on-site Coulomb interaction?
It turns out that an existence of the Fermi surface is not necessary for the superconducting phase.

We shall consider a new family of integrable models and show that
superconducting phase and strongly interacting Luttinger liquid states can
coexist at a high electron density or at small doping. This coexistence takes place at a
repulsive on-site Coulomb interaction the value of which is large than a
band width and depends on the value of the hard-core radius.
In the models \cite{EKS}, \cite{AAS}, \cite{DM} the
hoppings of single electrons on occupied states are forbidden, whereas the
energy of electron pair is finite. In our models we shall use the same hierarchy for the
parameters of the interactions, the constants
of interactions between single electrons are infinite and define a
hard-core radius, the energy of electron pair is finite. We shall
consider a new modification of a generalized one-dimensional Lai-Sutherland
model for a study of a competition between strongly interacting
Luttinger liquid state and superconducting phase. The model Hamiltonian contains
kinetic and interaction terms that combine those of the Hubbard model and
the Lai-Sutherland model. The model Hamiltonian includes two terms ${\cal H=H%
}_{hop}+{\cal H}_{int}$
\begin{eqnarray}
{\cal H}_{hop}&=&-t\sum_{<i,j>\sigma =\uparrow ,\downarrow }({\cal P}_{l}
c_{i \sigma}^{\dagger}c_{j\sigma}{\cal P}_{l} -
c_{i \sigma}^{\dagger}c_{j\sigma}n_{i -\sigma}n_{j -\sigma}), \\
{\cal H}_{int}&=&J\sum_{j=1}^{{\rm L}}\sum_{\sigma ,\sigma ^{\prime}=\uparrow ,\downarrow }
(c_{j \sigma}^{\dagger}c_{j\sigma^{\prime}}c_{j+1+l \sigma^{\prime}}^{\dagger}c_{j+1+l \sigma}
+ \nonumber \\
&&n_{j \sigma}n_{j+1+l \sigma^{\prime}})
+U\sum_{j=1}^{\rm L}\sum_{\sigma =\uparrow ,\downarrow } n_{j \sigma} n_{j - \sigma},
\end{eqnarray}
where $c_{j \sigma}^{\dagger}$ and $c_{j \sigma}$ are the creation and annihilation
operators of fermions with spin $\sigma$, $\sigma \in \{\uparrow
,\downarrow \}$,
{\rm L} is the total number of lattice sites, $<i,j>$ stands for neighboring sites,
the projector ${\cal P}_{l }$ forbids two single electrons at distances less than
or equal to {\it l} ( {\it l} is measured in units of the lattice
spacing parameter), $t$ is the hopping integral, $J$ is the constant of the exchange interaction. It is
important the ${\cal P}_{l}$ operator does not forbid doubly occupied
lattice sites, as it takes place in the so-called $U\rightarrow \infty $
Hubbard model or the t-J model. The last term in (2) is traditionally the
most important term for the Hubbard model, the on-site Coulomb repulsion
{\it U} separates the energies of single and paired electrons states. The
Hamiltonian ${\cal H}$ conserves not only the total number of electrons\
{\rm N} and also the number of single electrons with spin $\sigma $ ${\rm N}
_{1\sigma }=\sum_{j=1}^{L}n_{j \sigma}(1-n_{j -\sigma})$ and the number of
localized electron pairs ${\rm N}_{2}=\sum_{j=1}^{L}n_{j \uparrow} n_{j \downarrow},
{\rm N}=\sum_{\sigma }{\rm N}_{1\sigma }+2{\rm N}_{2}$. In the case {\it l}=0
and $J=0$  the
Hamiltonian (1), (2) is reduced to \cite{AAS} and for {\it l}=0, $U=\infty $
and $J=t$ to the Lai-Sutherland model \cite{LS}. For {\it l}$
>0$ the ${\cal P}_{l}$ operator is equivalent to additional two
particle interactions between single electrons $\sum_{r=1}^{l
}\sum_{j=1}^{L}U_{r} n_{1j}n_{1j+r}$, here
$n_{1j}=\sum_{\sigma =\uparrow \downarrow}n_{j \sigma}(1-n_{j -\sigma})$
with infinite $U_{r}$ parameters, according to (2) $U_{l +1}=-J$.
Using this representation we can conclude that the kinetic term of the
Hamiltonian (1) is a particle-hole invariant: indeed applying this
transformation $c_{j\sigma}^{\dagger}\Rightarrow c_{j \sigma},
c_{j\sigma}\Rightarrow c_{j \sigma}^{\dagger}$, to the Hamiltonian (1), (2) we obtain
${\cal H(}t,J,U)\Rightarrow {\cal H(}t,J,U)+U({\rm L}-{\rm N})$ . Due to a
particle-hole symmetry the phase diagram is symmetrical with respect to a
half filling.

We examine the exact ground state phase diagram for the
antiferromagnetic coupling $J=t$ (we chose the hopping integral equal to
unit then the coupling constants are dimensionless) and different values of
the hard-core radius. The results of calculations are compared with the ones
for $J=0$ - the simplest version of the model . Direct calculations show
that the model (1), (2) is an exactly solvable one by the Bethe ansatz
method and the set of the quasimomenta $\{k_{j}\}(j=1,2,...,{\rm N}_{1})$
satisfies the Bethe equations \cite{IK}
\begin{eqnarray}
\left( {\frac{{\lambda _{j}^{{}}-i/2}}{\lambda _{j}^{{}}+i/2}}\right) ^{{\rm %
L}- l {\rm N}_{1}} &=&(-1)^{N-1}\exp (-i l {\rm P})\prod_{i=1}^{{\rm %
M}}{\frac{{\lambda _{j}^{{}}-\lambda _{i}^{{}}-i}}{\lambda _{j}^{{}}-\lambda
_{i}^{{}}+i}}  \nonumber \\
&&\prod_{\alpha =1}^{{\rm M}}{\frac{{\lambda _{j}^{{}}-\chi _{\alpha
}^{{}}+i/2}}{\lambda _{j}^{{}}-\chi _{\alpha }^{{}}-i/2}},  \nonumber \\
\prod_{j=1}^{{\rm N}_{1}}{\frac{{\ \chi _{\alpha }^{{}}-\lambda _{j}^{{}}+i/2%
}}{\chi _{\alpha }^{{}}-\lambda _{j}^{{}}-i/2}} &=&-\prod_{\beta =1}^{{\rm M}%
}{\frac{{\chi _{\alpha }^{{}}-\chi _{\beta }^{{}}+i}}{\chi _{\alpha
}^{{}}-\chi _{\beta }^{{}}-i}},
\end{eqnarray}
where ${\rm P}=\sum_{j=1}^{{\rm N}_{1}}k_{j}$ is the momentum, ${\lambda }%
_{j}=\frac{1}{2}\tan \frac{k_{j}}{2}$ and ${\chi _{\alpha }^{{}}(\alpha
=1,2,...,}{\rm M}{)}$ are the 'charge' and 'spin' rapidities,{\rm \ }${\rm M}
${\rm \ }is the number of down spin single electrons.

The eigenvalues and the magnetization are given by

\begin{equation}
E=-2\sum_{j=1}^{{\rm N}_{1}}\cos k_{j}+U{\rm N}_{2},
\end{equation}

\begin{equation}
S^{z}=\frac{1}{2}\sum_{\sigma }{\rm N}_{1\sigma }-{\rm M}.
\end{equation}

Let us introduce the partial electron densities: ${\rm n}_{1}{\rm =N}_{1}%
{\rm /L}${\it \ } is the density of single carriers (${\rm N}%
_{1}=\sum_{\sigma =\uparrow, \downarrow}{\rm N}_{1\sigma }$),{\rm \ }
${\rm n}_{2}{\rm =N}_{2}{\rm /L}$ is the density of localized electron pairs. Clearly{\rm \ }
${\rm n=n}_{1}+2{\rm n}_{2}${\rm ,} here ${\rm n=N/L}$ is the total density of
electrons.

Since the Bethe equations (3) we can calculate exactly the ground state
phase diagram as a function of the electron density and an on-site
interaction for an arbitrary value of the hard-core radius. The density of
localized pairs ${\rm n}_{2}$ (or ${\rm n}_{1}$) can be calculated by
minimizing the ground state energy per site ${\cal E}=E/{\rm L}$ for a fixed
total density of electrons

\begin{equation}
{\cal E}=2{\rm n}_{1}-2\pi \int_{-Q}^{Q}d\Lambda a(\Lambda )\rho (\Lambda )+U%
{\rm n}_{2},
\end{equation}
where $a(\Lambda )=\frac{1}{2\pi }\frac{1}{\Lambda ^{2}+1/4}$.

In the thermodynamic limit the Bethe equations reduce to an integral
equation of the Fredholm type for the function of the distribution of $%
\lambda _{j}$ on the real axis
\begin{equation}
\rho (\Lambda )+\int_{-Q}^{Q}d\Lambda ^{\prime }K(\Lambda -\Lambda ^{\prime
})\rho (\Lambda ^{\prime })=(1-l{\rm n}_{1})a(\Lambda ),
\end{equation}
with the kernel being $K(\Lambda )=\int_{-\infty }^{\infty }\frac{d\omega }{%
2\pi }{\frac{\exp (-\mid \omega \mid )}{\exp \mid \omega \mid +1}\exp
(i\omega \Lambda )}$. The $\Lambda -$ Fermi level denoted as $Q$ controls
the band filling, the density of single electrons is defined by
\begin{equation}
{\rm n}_{1}=\int_{-Q}^{Q}d\Lambda \rho (\Lambda ).
\end{equation}
$Q=0$ corresponds to an empty subband of single carriers, ${\rm n}_{1}={\rm n%
}_{0}$ for $Q\rightarrow \infty $, here ${\rm n}_{0}=\frac{1}{l +3/2}$
is a 'half-filled' density. Equations (6)-(8) are a consequence of the real
solutions for 'charge' and 'spin' rapidities that describe the ground state
of the system in the absence of an external magnetic field. We calculate the
critical exponent $\Theta $ for a definition of the realization criterion
of strongly interacting Luttinger liquid. We remind the momentum
distribution function close to the Fermi momentum $k_{F}$ is determined by
the critical exponent $\Theta $
\begin{equation}
\langle n_{k}\rangle \simeq \langle n_{k_{F}}\rangle -{\rm const}%
|k-k_{F}|^{\Theta }{\rm sgn}(k-k_{F}),
\end{equation}
where $\Theta ={\frac{{1}}{\alpha }}\left( 1-{\frac{{\alpha }}{4}}\right)
^{2}$and $\alpha =2\zeta ^{2}(Q)$ is defined by the dressed charge $\zeta
(\Lambda )$ according to the following integral equation $\zeta (\Lambda
)+\int_{-Q}^{Q}d\Lambda ^{\prime }K(\Lambda -\Lambda ^{\prime })\zeta
(\Lambda ^{\prime })=1- l {\rm n}_{1}$. In the high electron density $%
{\rm n>2n}_{2+}{\rm n}_{c}$ (where ${\rm n}_{c}$ is solution of equation $%
\Theta ({\rm n}_{c})=1$ ) when the hard-core repulsive potential dominates
the behavior of fermions is described as strongly interacting Luttinger
liquid with $\Theta >1$ \cite{IK}.

%%%%%%%%%%%%%%%%%%%%%%%%%%%%%%%%%%%%%%%
\begin{figure}[tbp!]
\begin{center}
\includegraphics[width=0.45\textwidth,clip,bb=115 445 363 688]{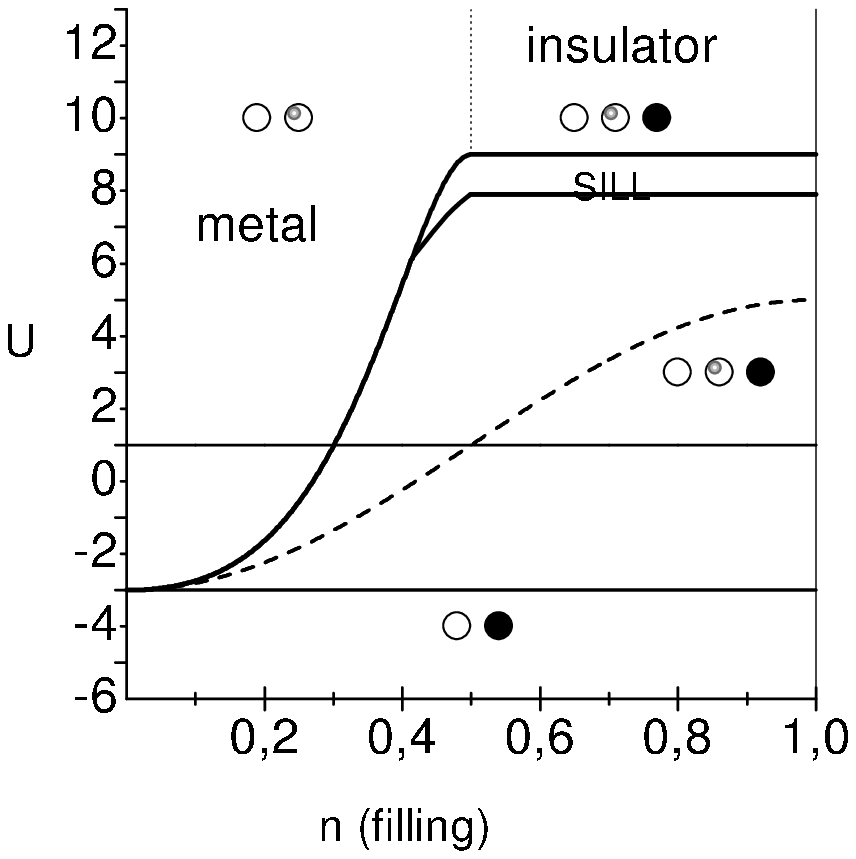}
\end{center}
\caption{\label{1}Ground state phase diagram in the case $J=0$ for $l=1$.
Dashed line corresponds to the model \protect\cite{AAS}
(or $l=0$), dotted lines separate metallic and insulator phases. The
area of a strongly interacting Luttinger liquid state is denoted as SILL. }
\end{figure}
%%%%%%%%%%%%%%%%%%%%%%%%%%%%%%%%%%%%%%%%

We have focused on the calculation of the exact ground state phase diagram
in the ${\rm n-}U$ plane for different values of the hard core radius or $l$.
First we consider peculiarities of behavior of the system using a
simple version of the Hamiltonian (1),(2) when $J=0$ and then its
transformation for $J=1$. Due to a particle-hole symmetry it is sufficient to
discuss the phase diagram for {\rm n}$\leq 1$. For $J=0$ the density of the
ground-state energy (6) can be defined analytically ${\cal E=}-2\frac{%
1-l {\rm n}_{1}}{\pi }\sin {\pi {\rm n}_{1} \overwithdelims() 1-l {\rm n}_{1}}
+\frac{1}{2}U({\rm n}-{\rm n}_{1})$, therefore a curve that separates a mixed
region is defined according to the following equation $U({\rm n}_{1})=4\frac{l}{\pi }\sin
%TCIMACRO{\QOVERD( ) {\pi {\rm n}_{1}}{1-\Delta {\rm n}_{1}}}
%BeginExpansion
{\pi {\rm n}_{1} \overwithdelims() 1-l {\rm n}_{1}}
-\frac{4}{1- l n_{1}}\cos
%TCIMACRO{\QOVERD( ) {\pi {\rm n}_{1}}{1- l {\rm n}_{1}}}
{\pi {\rm n}_{1} \overwithdelims() 1- l {\rm n}_{1}}
$. $U$ variates from $-4$ at {\rm n}$_{1}=0$ to $4(1+ l )$ at ${\rm n}
$. $U$ variates from $-4$ at {\rm n}$_{1}=0$ to $4(1+ l )$ at ${\rm n}
_{1}={\rm n}_{\max }=1/(1+l)$, hence a maximal value $U_{c}=4(1+l
)$. The value of ${\rm n}_{c}$ is equal to {\rm n}$_{c}=(1-\sqrt{6-4\sqrt{2}}%
)/l $ ({\rm n}$_{c}=0.414$ for $l=1,{\rm n}_{c}=0.207$ for $l =2$,
{\rm n}$_{c}=0.138$ for $l =3$). For $l=1$ the complete phase diagram
is shown in Fig.\ \ref{1}. The lower region ( for $U<-4$ ) is characterized by only
double occupied (solid circles) and empty (empty circles) sites so ${\rm n}%
_{1}=0$ and {\rm n}$_{2}=${\rm n}$/2$. For $-4<U<U({\rm n}_{1})$ we have a
mixed region, the ground state includes both finite densities of single
electrons (spheres with dot center) and localized electron pairs. These
phases are superconducting, since the two particle correlation function $%
<\eta _{i}^{\dagger }\eta _{j}>$ (here
$\eta_{j}^{\dagger }=c_{j \uparrow}^{\dagger}c_{j\downarrow}^{\dagger}$
) exhibits ODLRO \cite{EKS}, \cite{YY} i.e. $<\eta _{i}^{\dagger }
\eta_{j}> \nrightarrow 0$ for $\mid i-j\mid \rightarrow \infty $. At {\rm n}${\rm %
>}${\rm n}$_{c}$ a curve $U({\rm n}_{c})$ separates this phase on Luttinger
liquid state at $-4<U<U({\rm n}_{c})$ and strongly interacting Luttinger
liquid state at $U({\rm n}_{c})<U<U({\rm n}_{1})$.The strongly interacting Luttinger liquid state
is denoted as SILL in the figures. Note, that strongly
interacting Luttinger liquid state is realized at largest values of a
repulsive on-site Coulomb interaction and a high electron density. Comparing
the calculations for different {\it l} we can conclude that a hard-core
repulsive interaction increases a region of the coexistence of strongly
interacting Luttinger liquid and superconducting phase due to both a larger $%
U_{c}$ and smaller ${\rm n}_{c}$. For $U>U({\rm n}_{1})$ and $n<n_{max}$ the ground state coexists
of singly occupied and empty sites; dotted lines separate a
metallic phase (at ${\rm n}<{\rm n}_{\max }$) and an insulator phase (at ${\rm n\geq n%
}_{\max }$) with a gap $\Delta \varepsilon =U-U_{c}$.

\begin{figure}[tbp!]
\begin{center}
\includegraphics[width=0.45\textwidth,clip,bb=120 445 363 688]{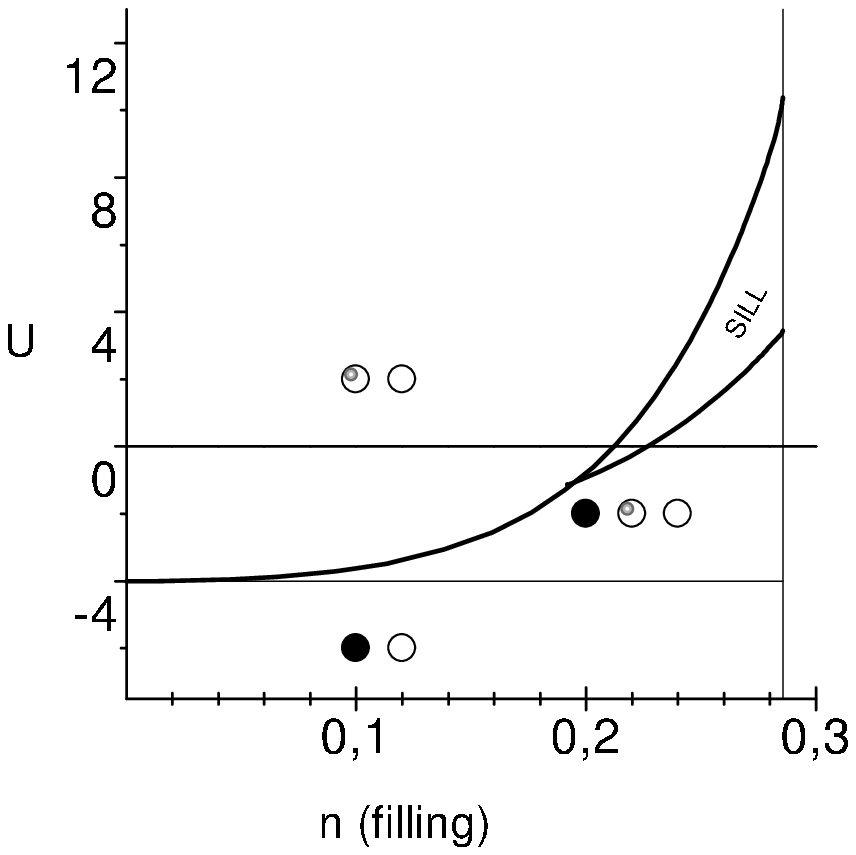}
\end{center}
\caption{\label{2}Ground state phase diagram in the case $J=1$ and $l=2$ - similar to that for
Fig.\ \ref{1}.}
\end{figure}

An exact solution of the problem enables to study the role of the exchange
interaction on the behavior of a strongly interacted electron system. Let us
consider a transformation of the exact ground-state phase diagram for $J=1$,
we restrict our consideration the case ${\rm n\leq n}_{0}$. $\Theta $
increases monotonically from $\frac{1}{8}$ to $\frac{(3+2l )^{2}}{12}[1-%
\frac{3}{(3+2 l )^{2}}]^{2}$ with the ${\rm n}_{1}$ density. We should
solve equation $\Theta ({\rm n}_{c})=1$ numerically calculating the dressed
charge as a function of the electron density ${\rm n}_{1}$ for arbitrary $%
l$; for example ${\rm n}_{c}=0.348$ for $l=1$, ${\rm n}%
_{c}=0.192 $ for $l=2$, ${\rm n}_{c}=0.131$ for $l=3$, ${\rm n}%
_{c}=0.1$ for $l =4$. According to the numerical results obtained the
critical density ${\rm n}_{c}$ is less than ${\rm n}_{0}$.For $l=2$ the ground state
phase diagram is given in Fig.\ \ref{2}. All electron states: empty, single
occupied and doubly occupied sites are presented simultaneously in a mixed
region (a closed region in figure 2). For ${\rm n}_{c}<{\rm n}<{\rm n}_{0}$
two branches of curves separate Luttinger liquid
state and strongly interacting Luttinger liquid, that is realized between
these branches. Comparing the phase diagrams for $J=1$ (Fig.\ \ref{2}) and
$J=0$ (Fig.\ \ref{1}) calculated for the same value of $l$ we can
conclude that the exchange interaction decreases the region of the
coexistence of the superconducting phase and strongly interacting Luttinger
liquid state (${\rm n}_{c}$ and $U_{c}$ decrease slightly).
$U_{c}$ increases with an increasing of hard-core radius.

%%%%%%%%%%%%%%%%%%%%%%%%%%%%%%%%%%%%%%%%%%
%%%%%%%%%%%%%%%%%%%%%%%%%%%%%%%%%%%%%%%%%%%

In summary, we have presented a soluble generalization of the Lai-Sutherland
model, having the nontrivial Luttinger liquid behavior. The exact solution
was obtained by means of the nested Bethe Ansatz. We have
derived the exact ground-state phase diagram; the latter exhibits an unusual phase
state in which superconducting phase and strongly interacting
Luttinger liquid state coexist. This phase is realized at high electron density and
positive values of the on-site Coulomb interaction. The maximum
critical value $U_{c}$ realized in the model is higher than that of all other
exactly solvable models \cite{AAS},\cite{DM}. This is important because higher values of $U_{c}$
expands the region of coexistence of superconducting phase and strongly
interacting Luttinger liquid state. It has been shown that the presence of the Fermi level is
not necessary for realization of superconducting phase.
The results of calculations of one dimensional models do not allow direct application
to the real 2D and 3D systems. Nevertheless one can assume that real high-T$_{c}$ superconductors
belong to the family of strongly interacting Luttinger liquid described
above.

%\acknowledgments
\begin{acknowledgments}
IK would like to thank Prof. E.D. Belokolos for valuable discussions and wishes to thank
the support of the Visitor Program of the Max-Planck-Institut f\"{u}r Physik Komplexer Systeme, Dresden, Germany.
This work is supported under RFFR Grants No 00-03-32981, No 00-15-97334 and the STCU-2354 project.
\end{acknowledgments}

%%%\begin{references}


\begin{thebibliography}{99}
\bibitem{EKS}  F.H.L. Essler, V.E. Korepin, and K. Schoutens, Phys. Rev.
Lett. {\bf 68}, 2960 (1992); {\it ibid.} {\bf 70}, 73 (1993);
An.A. Ovchinnikov, Mod. Phys. Lett. B {\bf 7}, 1397 (1993).

\bibitem{AAS}  L. Arrachea and A.A. Aligia, Phys. Rev. Lett. {\bf 73}, 2240
(1994); A. Schadschneider, Phys. Rev. B {\bf 51}, 10386 (1995).

\bibitem{DM}  F. Dolcini and A. Montorsi, Phys. Rev. B {\bf 63}, R121103
(2001); F. Dolcini and A. Montorsi, Nucl.Phys. B {\bf 592}, 563 (2001).

\bibitem{IK}  I.N. Karnaukhov, Europhys. Lett. {\bf 52}, 571 (2000); I.N.
Karnaukhov and N. Andrei, J. Phys.: Condens. Matter {\bf 13}, L891 (2001);
I.N. Karnaukhov and A.A. Ovchinnikov, Europhys. Lett. {\bf 57}, 540 (2002).

\bibitem{LS}  C.K. Lai, J.Math. Phys. {\bf 15},1675 (1974); Bill Sutherland,
Phys. Rev. B {\bf 12, }3795 (1975).

\bibitem{YY}  C.N. Yang, Phys. Rev. Lett. {\bf 63}, 2144 (1989); C.N. Yang
and S. Zhang, Mod.Phys.Lett. B{\bf \ 4, }759 (1990).


%\end{references}  
\end{thebibliography}
\end{document}